\documentclass[]{JFM-FLM_Au}
\usepackage{algorithmic}
\usepackage{threeparttable}
\usepackage{subcaption}
\usepackage{siunitx}
\usepackage{orcidlink}
\usepackage[abs]{overpic}
\usepackage[T1]{fontenc}
\usepackage{graphicx}

\lefttitle{F. Xing et~al.}
\righttitle{Dynamic Interface Modeling via Constrained PINNs}

\title{Modeling Dynamic Gas–Liquid Interfaces in Underwater Explosions Using Interval-Constrained Physics-Informed Neural Networks}

\author{
Fulin Xing\aff{1}\footnotemark[1],
Junjie Li\aff{1}\thanks{These authors contributed equally},
Ze Tao\aff{1}\footnotemark[1],
Fujun Liu\aff{1}\corresp{ Email address for correspondence: \email{fjliu@cust.edu.cn}}
Yong Tan\aff{2}}

\affiliation{\aff{1}{Nanophotonics and Biophotonics Key Laboratory of Jilin Province, School of Physics, Changchun University of Science and Technology},
               {Changchun},
                {130022},
                {PR China}; \aff{2}{Province Key Laboratory of Spectral Exploration Science and Technology, School of Physics, Changchun University of Science and Technology},
               {Changchun},
                {130022},
                {PR China}
}

\begin{document}
\maketitle

\begin{abstract}
Underwater explosion modeling faces a critical challenge of simultaneously resolving shock waves and gas-liquid interfaces, as traditional methods struggle to balance accuracy and computational efficiency. To address this, we develop a physics-informed neural network (PINN) framework featuring a dual-network architecture, that one network learns flow-field variables (pressure, density, velocity) from simulation data, while another network tracks the gas-liquid interface despite lacking direct numerical solutions. Crucially, we introduce an interval-constraint training strategy that penalizes interface deviations beyond grid spacing limits, paired with a physics-preserving linear mapping of 1D spherical Euler equations to ensure consistency. Our results show that this approach accurately reconstructs spatiotemporal fields from coarse-grid data, achieving superior computational efficiency over conventional CFD—enabling rapid, mesh-free blast-load analysis for near/far-field scenarios and extensibility to higher-dimensional problems.
\end{abstract}



\section{Introduction}
Underwater explosions (UNDEX) exhibit significantly greater destructive power and damage radius compared to air explosions, primarily due to water's efficient shock wave transmission \citep{cranz1925lehrbuch,hopkinson1915british}. These phenomena involve complex gas-liquid interactions characterized by highly nonlinear pressure propagation \citep{jia2022nonlinear} and intricate interface dynamics \citep{zhang2023unified}. Beyond military applications like ship damage assessment \citep{he2023damage} and torpedo design \citep{zhang2023review}, UNDEX research holds substantial value for hydraulic engineering \citep{yang2023experimental,wang2020blast,huang2024damage,zhou2024dynamic}. The field primarily focuses on two interdependent aspects: shockwave propagation and subsequent bubble dynamics. Accurate modeling requires integrating both components - where high-pressure gas drives initial perturbations in water, while bubble motion is constrained by fluid boundaries. This integration presents a fundamental challenge: simultaneously solving Euler conservation equations for the full flow field while precisely tracking the evolving gas-liquid interface to determine domain state equations.

Numerical approaches to this challenge must satisfy stringent robustness requirements. The Arbitrary Lagrangian-Eulerian (ALE) method \citep{barras2012numerical,wu2024study} balances Eulerian stability and Lagrangian efficiency through controlled mesh stiffness and particle-based interface tracking. While \citet{lohner2023arbitrary} improved stability using hyperbolic tangent interface capturing, ALE methods still struggle with mesh distortion during bubble pulsation and jet formation. Alternatively, the Coupled Eulerian-Lagrangian (CEL) method \citep{meng2019study,nguyen2020application} employs Eulerian grids for fluids and Lagrangian grids for structures, enabling effective damage assessment \citep{bardiani2025underwater} but sacrificing interface precision through diffuse Eulerian volume fraction (EVF) techniques. These mature methods, widely implemented in commercial software \citep{tran2021composite}, are complemented by research-focused approaches: Eulerian Finite Element Methods (EFEM) \citep{he2020prolonged,liu2024pressure} using computationally expensive Volume-of-Fluid (VOF) interface tracking, and Ghost Fluid Methods (GFM) \citep{jin2025correction} that suffer from boundary oscillations. Although subsequent improvements like the Practical GFM \citep{Xu_Feng_Liu_2016} and Modified GFM \citep{feng2020modified} have reduced numerical errors, and \citet{si2021framework} minimized interpolation errors, simultaneously resolving sharp interfaces and shock fronts remains challenging \citep{shu2023underwater}. Collectively, these traditional methods face persistent accuracy-efficiency tradeoffs in UNDEX modeling.

Physics-informed neural networks (PINNs) \citep{raissi2020hidden} offer a paradigm shift by embedding physical laws directly into neural network architectures through PDE-based loss functions. Unlike conventional CFD, PINNs eliminate explicit meshing and time-marching schemes through their black-box training approach. Demonstrating particular promise in fluid dynamics, PINNs have successfully modeled arterial blood flow and cylinder wakes \citep{raissi2020hidden}, with  \citet{mishra2022estimates} enhancing reliability through L2-error norms and  \citet{cai2021physics} scaling the framework to high-dimensional problems. These advances suggest significant potential for UNDEX applications.

Building on these developments, we propose a novel interval-constrained PINN method that integrates weak supervision signals with physical conservation laws for moving boundary modeling. Our key innovation lies in constraining the interface network using coarse CFD mesh intervals - penalizing predictions outside these bounds while explicitly incorporating interface positions into the Euler equation training. The approach features three core components: (1) a dual-network architecture separating flow-field and interface predictions to prevent unconverged interfaces from corrupting flow solutions; (2) a geometric mapping that transforms dynamic domains into fixed reference coordinates while preserving spherical source terms; and (3) rigorous enforcement of interface continuity conditions through pressure-balance and kinematic constraints. Numerical validation demonstrates simultaneous prediction of flow fields and interface positions, offering a new paradigm for blast-load evaluation that captures both near-field nonlinearities and far-field shock propagation with superior computational efficiency compared to traditional CFD. The paper is organized as follows: Section~\ref{sec:2} derives the governing equations and interface conditions, Section~\ref{sec:3} details the interval-constrained PINN methodology, Section~\ref{sec:4} presents Training results, Section~\ref{sec:5} validates assessment capability for UNDEX, and Section~\ref{sec:6} concludes with future research directions.

\section{Theory}\label{sec:2}
In this study, we investigates the fluid dynamics of an UNDEX. We consider a spherical TNT charge detonating in an unbounded fluid domain, neglecting external boundaries to isolate the fundamental physics of the explosion. To focus on the dominant mechanisms, we make three key simplifying assumptions: 

\textit{a) gravity is neglected, eliminating bubble buoyancy effects}; 

\textit{b) viscosity is ignored, removing shear stresses between materials}; 

\textit{c) heat conduction is omitted, justified by the adiabatic nature of the transient event}. 

\noindent These assumptions reduce the problem to a one-dimensional spherically symmetric system, where all flow quantities (pressure, density, velocity) depend solely on the radial distance $r$ from the charge center and time $t$. Consequently, the governing equations simplify to the one-dimensional spherically symmetric Euler equations, capturing the essential compressible flow dynamics while maintaining computational tractability.

\subsection{Spherically Symmetric One-Dimensional Euler Equations}

\begin{figure}
    \centering
	\includegraphics[width=0.6\textwidth, angle=0]{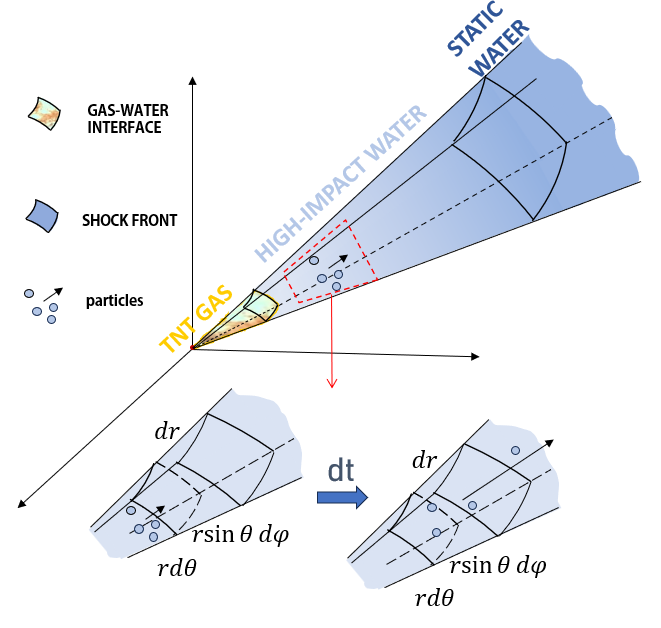}	
	\caption{ Spherical explosion model. \textit{(Upper)} Schematic showing material distribution and interface positions. \textit{(Bottom)} Geometric relationship of a fixed spatial volume element ($dV$): at time $t$, water particles occupy the lower surface; during a time interval $dt$, particles flow into the element while others exit through the upper surface.} 
	\label{fig1}
\end{figure}

We derive the continuity equation for mass conservation by analyzing a fixed control volume in a spherical coordinate system. The fundamental principle requires that the rate of mass change within the control volume must balance the net mass flux through its boundaries.

Consider a differential volume element in spherical coordinates (figure~1), where the contained mass $dm$ is given by:
\begin{equation}
dm = \rho r^{2}\sin\theta , dr , d\theta , d\varphi
\end{equation}
with $\rho$ representing the local density field. For this fixed spatial location, the temporal rate of mass change becomes:
\begin{equation}
\frac{\partial}{\partial t}(dm) = \frac{\partial \rho}{\partial t} r^{2}\sin\theta , dr , d\theta , d\varphi \label{2}
\end{equation}

The mass flux analysis simplifies due to purely radial particle velocity. The incoming flux through the lower surface at radius $r$ is:
\begin{equation}
Q_m(r) = \rho u r^2 \sin \theta , d\theta , d\varphi
\end{equation}
where $u$ denotes the radial velocity component. Applying a first-order Taylor expansion gives the net flux difference across the differential element:
\begin{equation}
Q_m(r+dr) - Q_m(r) = \left(\frac{\partial (\rho u)}{\partial r} + \frac{2\rho u}{r}\right) r^2 \sin \theta , dr , d\theta , d\varphi \label{4}
\end{equation}

Enforcing mass conservation by equating \eqref{2} and \eqref{4} yields the spherical continuity equation:
\begin{equation}
\frac{\partial\rho}{\partial t} + \frac{\partial(\rho u)}{\partial r} = -\frac{2u}{r}\rho \label{3}
\end{equation}

The pressure differential across the volume element generates a net force that drives fluid motion. Applying Newton's second law to the differential element yields:
\begin{equation}
[p(r+dr) - p(r)]dS = \frac{\partial p}{\partial r}dr , dS = -\rho \frac{du}{dt}dV \label{6}
\end{equation}
where $p(r)$ and $p(r+dr)$ represent pressures at the lower and upper surfaces respectively, with $dS$ and $dV$ denoting differential surface area and volume. The material derivative of velocity expands to:
\begin{equation}
\rho \frac{du}{dt} = \rho\frac{\partial u}{\partial t} + \rho u\frac{\partial u}{\partial r} \label{7}
\end{equation}

Combining the continuity equation \eqref{3} with momentum relations \eqref{6}-\eqref{7} produces the spherical momentum conservation equation:
\begin{equation}
\frac{\partial(\rho u)}{\partial t} + \frac{\partial(\rho u^{2}+p)}{\partial r} = -\frac{2}{r}\rho u^{2}
\end{equation}

Energy conservation follows analogous considerations, balancing a) temporal energy change within the volume, energy flux through boundaries and pressure work performed on the fluid

This yields the energy equation:
\begin{equation}
\frac{\partial E}{\partial t} + \frac{\partial u(E+p)}{\partial r} = -\frac{2u}{r}(E+p)
\end{equation}
where $E \equiv \rho e + \frac{1}{2}\rho u^2$ represents the total energy density, combining specific internal energy $e$ and kinetic energy components. 

The complete set of conservation equations for the flow field has been derived, and a critical distinction from their Cartesian counterparts is the presence of geometric source terms $S$ in the spherical formulation. These terms take the form $\frac{2u}{r}$ multiplied by the respective field variable, representing the physical effect of geometric spreading as particles move radially outward. This phenomenon becomes particularly significant in the near-field region, where radial distances are small and the $\frac{1}{r}$ dependence leads to a stronger influence.
\subsection{Linear Mapping of Euler Equations}
The system of equations is closed through three region-specific equations of state, as illustrated in figure \ref{fig1}:

\begin{equation}
p(r,t)=
\begin{cases}
p_g, & \text{gas phase } (r < R) \\
p_l, & \text{liquid phase } (R\leqq r\leqq Rs) \\
0, & \text{ambient water } (Rs<r)
\end{cases}
\end{equation}

The gas-phase pressure $p_g$ for TNT products follows the JWL equation of state:
\begin{equation}
p_g = A\exp\left(-R_1 \frac{R^3}{R_c^3}\right) + B\exp\left(-R_2 \frac{R^3}{R_c^3}\right) + C\left( \frac{R}{R_c}\right)^{-3(\omega+1)}
\end{equation}
where $R$ represents the evolving gas-liquid interface position (the sole unknown), and other parameters are known constants \citep{zhang2008numerical}. For the liquid phase, we employ the Tait equation of state \citep{liu2018investigation}:
\begin{equation}
p_l = \rho e(\gamma -1) - \gamma P_w
\end{equation}
with $\gamma$ and $P_w$ as material constants. The shock front $Rs$ demarcates shocked and ambient water regions.

To focus computational efforts on the physically significant domain between interfaces $R$ and $Rs$, we introduce a dimensionless coordinate transformation:
\begin{equation}
\zeta = (r-R)/L, \quad  L = Rs-R
\end{equation}

This mapping yields the transformed Euler equations:

\begin{equation}
\begin{cases}
& \frac{\hat\partial \rho}{\hat\partial t} + \frac{\partial}{\partial \zeta} (\hat{u}\rho)
= \rho \left( \frac{-2u}{\zeta L + R} - \frac{1}{L}\frac{\hat\partial L}{\hat\partial t} \right) \\
& \frac{\hat\partial (\rho u)}{\hat\partial t} + \frac{\partial}{\partial \zeta} \left( \hat{u}\rho u + \frac{p}{L} \right)
= \rho u \left( \frac{-2u}{\zeta L + R} - \frac{1}{L}\frac{\hat\partial L}{\hat\partial t} \right) \\
& \frac{\hat\partial E}{\hat\partial t} + \frac{\partial}{\partial \zeta} \left( \hat{u}E + \frac{u p}{L} \right)
= -\frac{2u (E + P)}{\zeta L + R} - \frac{E}{L}\frac{\hat\partial L}{\hat\partial t}
\end{cases}
\end{equation}
where $\hat{u}$ denotes velocity in $\zeta$-coordinates and $\frac{\hat\partial}{\hat\partial t}$ indicates time derivatives at fixed $\zeta$.See Appendix~\ref{appA} for detailed derivation.

The interface position $R(t)$ plays a dual role: it appears explicitly in both the conservation equations and equations of state, ensuring physical consistency, while also serving as the primary disturbance source from the explosive. In contrast, the shock front $Rs(t)$ primarily defines the computational boundary, whose exact position has negligible impact on solution accuracy when sufficiently large. Consequently, our analysis treats $Rs$ as predetermined while focusing computational resolution on accurately capturing $R(t)$ dynamics.

Following the framework established by \citet{zhang2023unified}, we define the initial time as $\epsilon$ (a small finite time after shockwave emergence). The initial flow field conditions derive from shock jump relations:
\begin{equation}
\left\{
\begin{array}{l}
\rho_{\mathrm{initial}}
= \displaystyle\frac{\rho_{\infty}(\gamma+1)\,M^2}{(\gamma-1)\,M^2 + 2}\\[1.5ex]
u_{\mathrm{initial}}
= \displaystyle\frac{2\,C_{\infty}}{\gamma+1}\,\frac{M^2 - 1}{M}\,
\end{array}
\right.
\end{equation}
where $M$ represents the shock Mach number, and $C_{\infty}$, $\rho_{\infty}$ denote ambient sound speed and density respectively. Interface positions initialize as:
\begin{equation}
\begin{cases}
R_{\text{initial}} = R_c + \epsilon u_{\text{initial}} \\
Rs_{\text{initial}} = R_c + M C_\infty \epsilon
\end{cases}
\end{equation}

Boundary conditions enforce physical constraints:
\begin{equation}
\begin{cases}
p_{\text{left}} = p_g \\
u_{\text{left}} = \dot{R}
\end{cases}, \quad
\begin{cases}
\rho_{\text{right}} = \rho_\infty \\
u_{\text{right}} = 0
\end{cases}
\end{equation}

These conditions maintain conservation law compliance, requiring smooth continuity of $u$, $\rho$, and $p$ throughout the domain. The velocity boundary condition $u_{\text{left}} = \dot{R}$ particularly ensures kinematic consistency at the gas-liquid interface. Our subsequent PINN implementation will simultaneously solve for both the global flow field and the continuously evolving interface $R(t)$.

\begin{table}
\centering
\begin{tabular}{clccccclcccl}
\textbf{parameter }                          & $P_w$                  & \multicolumn{1}{l}{$\gamma$} & \multicolumn{1}{l}{$\rho_0$} & \multicolumn{1}{l}{$c_\infty$} & $A$                        & $B$                        & $C$                    & \multicolumn{1}{l}{$R_1$} & \multicolumn{1}{l}{$R_2$} & \multicolumn{1}{l}{$\omega$} & $R_c$                  \\ \hline
\textbf{Water}                              & 330.9                  & 7.15                         & 1000                         & 1450                           & --                         & --                         & \multicolumn{1}{c}{--} & --                        & --                        & --                           & \multicolumn{1}{c}{--} \\
\multicolumn{1}{l}{\textbf{TNT productions}} & \multicolumn{1}{c}{--} & --                           & 1630                         & --                             & \multicolumn{1}{l}{373.77} & \multicolumn{1}{l}{3.7471} & 0.74                   & 4.15                      & 0.9                       & 0.35                         & 0.5478       \\ \hline         
\end{tabular}
\begin{tablenotes}
\item[] $P_w$: water pressure constant [MPa]; $\gamma$: adiabatic index; $\rho_0$: density at rest [kg/m$^3$]; $c_\infty$: ambient sound speed in water [m/s]. Parameters $A$, $B$, and $C$ are constants in the JWL equation of state [GPa];  $R_1$, $R_2$: dimensionless exponents in the JWL equation; $\omega$: Grüneisen coefficient; $R_c$: charge radius [cm].
\end{tablenotes}
\caption{Physical parameters for water and TNT productions}
\label{tab:water_tnt_compact}
\end{table}

\section{Methods}\label{sec:3}
To address the computational complexity of underwater explosion (UNDEX) modeling, we strategically decompose the problem into two distinct regimes:

a) Far-field analysis: Simulates shock wave propagation through the water medium, where nonlinear effects are moderate but long-range accuracy is crucial

b) Initial phase analysis: Focuses on the highly transient evolution immediately following detonation, including
shockwave formation, early-time propagation, and gas-liquid interface motion within a confined domain;

This dual-approach methodology enables targeted treatment of each regime's unique physical characteristics while maintaining computational tractability.

\subsection{Configurations and Parameters in Simulation}
\begin{figure}[htbp]
\begin{subfigure}[t]{1.2\textwidth}
 \includegraphics[width=0.8\textwidth]{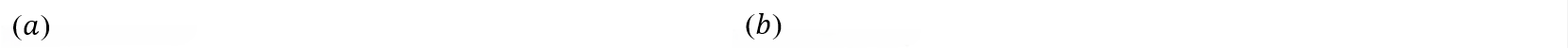}   
\end{subfigure}
\flushleft
\begin{subfigure}[t]{0.3\textwidth}
\includegraphics[width=2.7in]{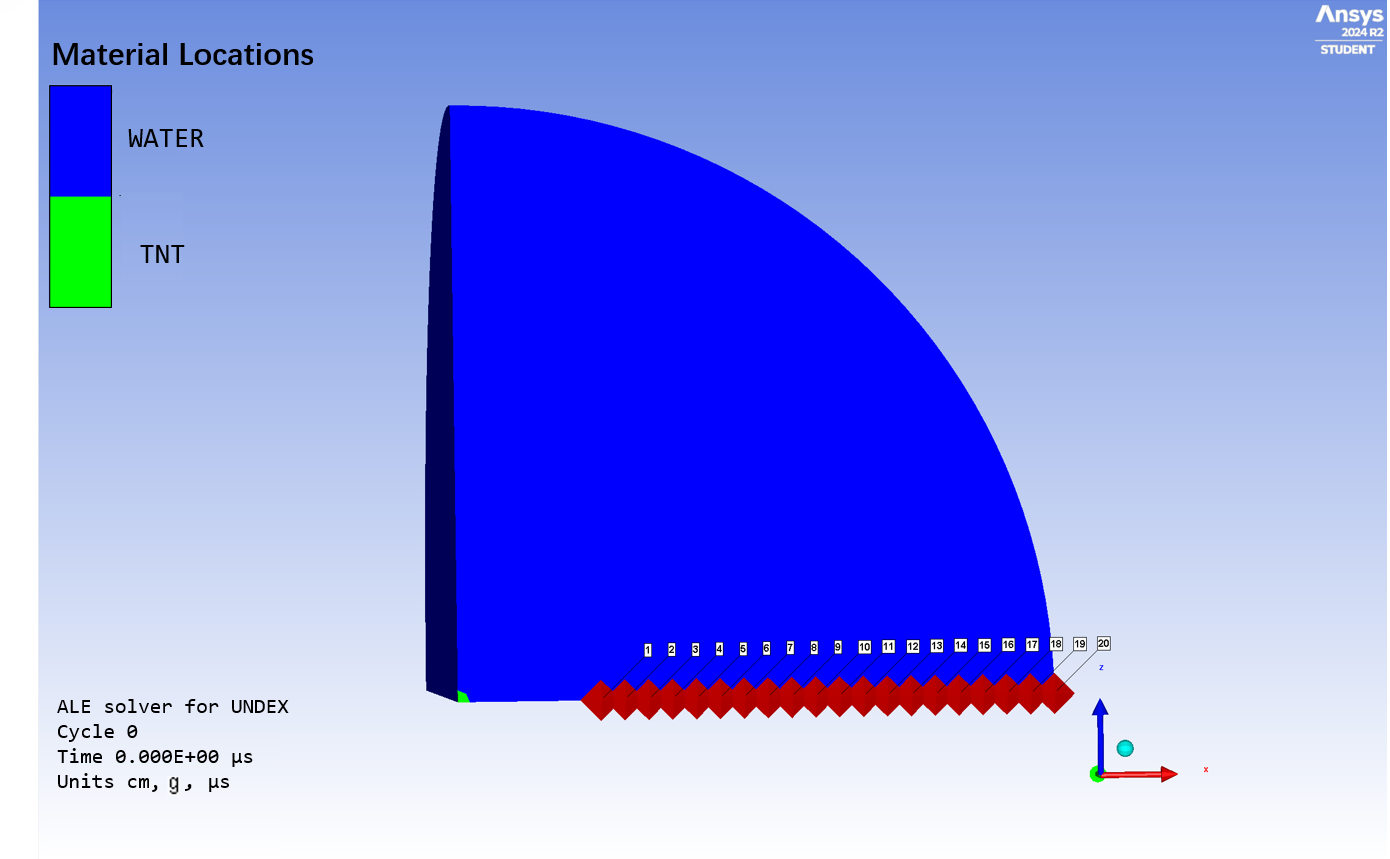}
\end{subfigure}
\hspace{20mm}
\begin{subfigure}[t]{0.3\textwidth}
\includegraphics[width=2.83in]{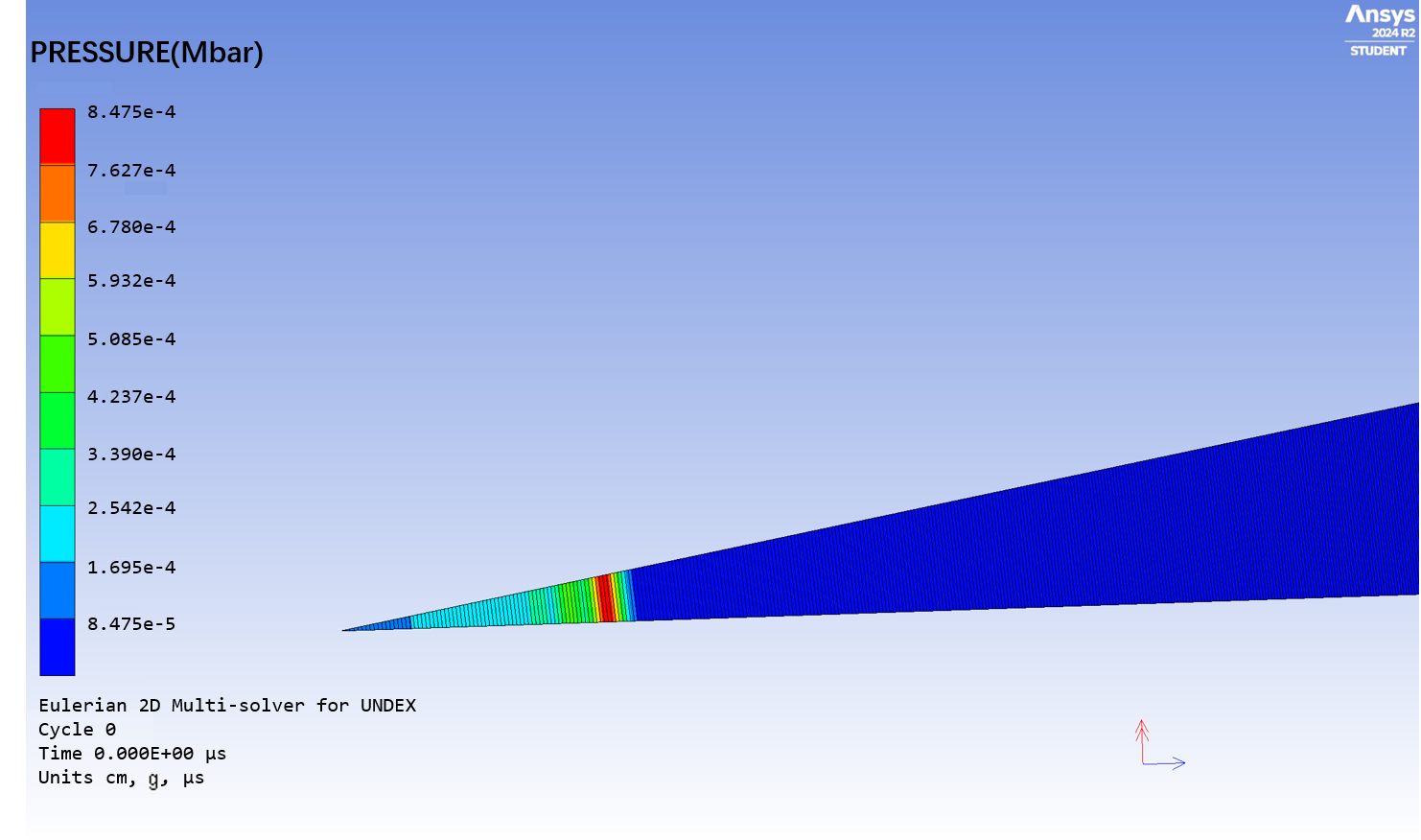}
\end{subfigure}
\caption{Simulation cases for underwater explosion analysis. \textbf{($a$)} Far-field shock propagation (Case 1) with observation points (red diamonds) positioned to avoid mesh distortion effects. \textbf{($b$)}  Initial phase pressure distribution (Case 2) showing complex gas-liquid interface dynamics during detonation.}
\label{fig2}
\end{figure}

All simulation data for this study were generated using the ANSYS Autodyn module. The computational setup was designed to capture both the explosive dynamics and fluid response with appropriate resolution as shown in figure \ref{fig2}. 

\textit{Case 1 Configuration:} We implemented a three-dimensional quarter-spherical domain using an Arbitrary Lagrangian-Eulerian (ALE) solver, discretized into a $40 \times 40 \times 40$ uniform grid with 1 cm cubic cells to achieve balanced resolution and computational efficiency. The material distribution consisted of TNT explosive within the 0-0.8 cm radius region surrounded by high-shock-response water, with rigid wall boundary conditions applied at the outer limits. The simulation ran for 105 $\mu$s with a 0.1 $\mu$s timestep, monitoring fluid dynamics through 20 equally spaced observation points along the x-axis (20-40 cm range). This configuration generated time histories of velocity ($u$), density ($\rho$), and pressure ($p$) that, after post-processing, yielded approximately 10,000 data points per variable for far-field shock wave analysis.

\textit{Case 2 Configuration:} For the initial phase of UNDEX, defined as the period during which the shock wave remains within ten times the instantaneous bubble radius, we employed a two-dimensional wedge-shaped domain (80 cm length) using an Eulerian 2D Multi-solver with finer 0.1 cm resolution cells. The first five cells contained TNT explosive within an extended computational domain designed to prevent wave reflection interference. With a 0.01 $\mu$s timestep over 150 $\mu$s duration, we recorded data from the first 300 grid points, producing approximately 110,000 data points per variable after processing. This high-resolution setup specifically captured the complex nonlinear interactions and interface motions characteristic of initial-phase UNDEX phenomena.

\subsection{Case 1: Disturbance Propagation in Stationary Water}

We model the physical process of a manually defined disturbance propagating through a stationary water domain by treating the first observation point from the simulation data as the disturbance source. The boundary condition is implemented as a linear interpolation of this point's physical field over time. The computational domain $\Omega_f$ is constructed as a fixed rectangular space-time coordinate system, where the horizontal axis represents radial position $r$ relative to the observation point and the vertical axis represents time $t$. Within this domain:

a) the left boundary receives disturbance input;

b) initial and right boundary conditions are set to static water states;

c) the solution $\mathbf{u}_f(x,t)$ satisfies the 1D spherical Euler equations

The PINN is trained using a composite loss function:
\begin{equation}
L(\theta)=\lambda_d L_{d}(\theta)+\lambda_{eq} L_{eq}(\theta)+\lambda_0 L_0 (\theta)+\lambda_b L_b(\theta)    
\end{equation}
where $\lambda_d$, $\lambda_{eq}$, $\lambda_0$, and $\lambda_b$ (all set to 0.5) weight the data, equation, initial condition, and boundary condition losses, respectively. The equation loss components, computed via automatic differentiation, enforce conservation laws:

\begin{align}
L_{eq1} &= \frac{1}{n_{eq}} \sum_{i=1}^{n_{eq}} \left\| \frac{\partial \rho_i}{\partial t} + \frac{\partial (u_i \rho_i)}{\partial x} +\frac{2u_i \rho_i}{x+x_0}\right\|^2 \\
L_{eq2} &= \frac{1}{n_{eq}} \sum_{i=1}^{n_{eq}} \left\| \frac{\partial (\rho_i u_i)}{\partial t} + \frac{\partial (u_i^2 \rho_i + p_i)}{\partial x}+\frac{2u_i^2 \rho_i}{x+x_0} \right\|^2 \\
L_{eq3} &= \frac{1}{n_{eq}} \sum_{i=1}^{n_{eq}} \left\| \frac{\partial E_i}{\partial t} + \frac{\partial [u_i(E_i + p_i)]}{\partial x}+\frac{2u_i(E_i + p_i)}{x+x_0} \right\|^2 \\
L_{eq} &= L_{eq1} + L_{eq2} + L_{eq3}
\end{align}
where $E_i \equiv E(x_i,t_i) = \frac{p_i+\gamma P_w}{\gamma -1} +\frac{1}{2}\rho_i u_i^2$ is computed from the equation of state, and $x_0$ marks the disturbance source location.

The remaining loss terms incorporate known physical constraints:
\begin{align}
L_d &= \frac{1}{n_d} \sum_{i=1}^{n_d} \|\mathbf{u}(x_i, t_i) - \mathbf{u}_d(x_i, t_i)\|^2 \quad \text{(Data)} \\
L_0 &= \frac{1}{n_0} \sum_{i=1}^{n_0} \|\mathbf{u}(x_i, 0) - \mathbf{u}_0(x_i)\|^2 \quad \text{(Initial)} \\
L_b &= \frac{1}{n_b} \sum_{i=1}^{n_b} \|\mathbf{u}(x_i, t_i) - \mathbf{u}_b(x_i, t_i)\|^2 \quad \text{(Boundary)}
\end{align}

The data points $n_d$ correspond to standard grid sampling, while $n_{eq}$, $n_0$, and $n_b$ employ Latin hypercube sampling for uniform domain coverage \citep{rao2020physics}. This approach ensures comprehensive constraint enforcement throughout the computational domain.

\begin{figure}
	\centering 
	\includegraphics[width=0.8\textwidth, angle=0]{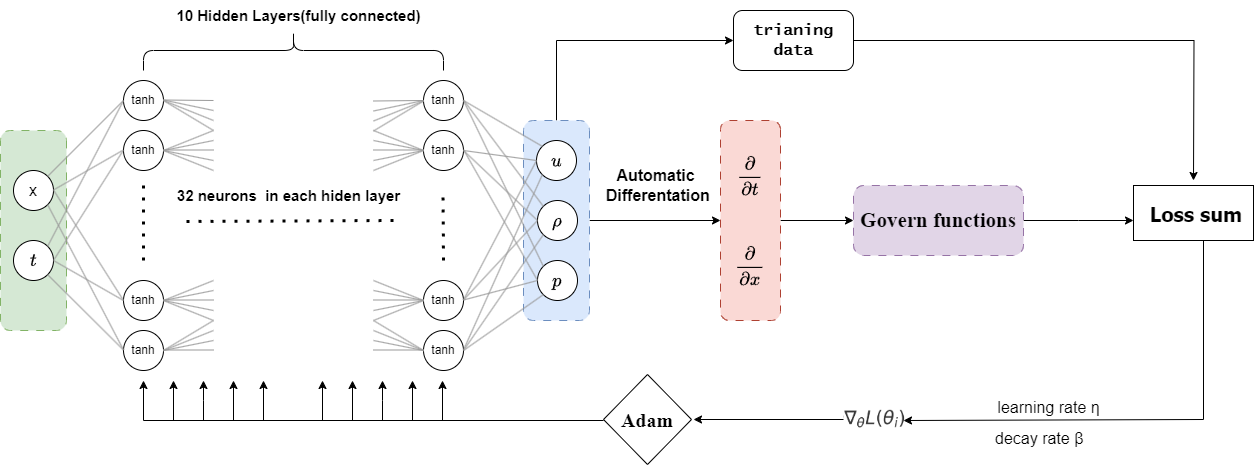}
	\caption{PINN network architecture for Case 1} 
	\label{fig3}
\end{figure}

The computational framework, illustrated in figure~\ref{fig3}, processes spatial and temporal coordinates through a fully connected neural network with tanh activation functions \citep{sharma2017activation} to predict the velocity ($u$), density ($\rho$), and pressure ($p$) fields. During training, the composite loss function - incorporating both data fidelity terms and PDE residuals - propagates gradient information of the network parameters $\theta$ to the Adam optimizer \citep{kingma2014adam}. This optimization algorithm dynamically adjusts the search direction by combining current gradients with historical momentum estimates, then updates the weights through a controlled step size determined by the learning rate $\eta$.

\subsection{Case 2: Dual-Network Architecture for Interface and Flow Field Modeling}

\begin{figure}[t]
	\centering 
\includegraphics[width=0.8\textwidth]{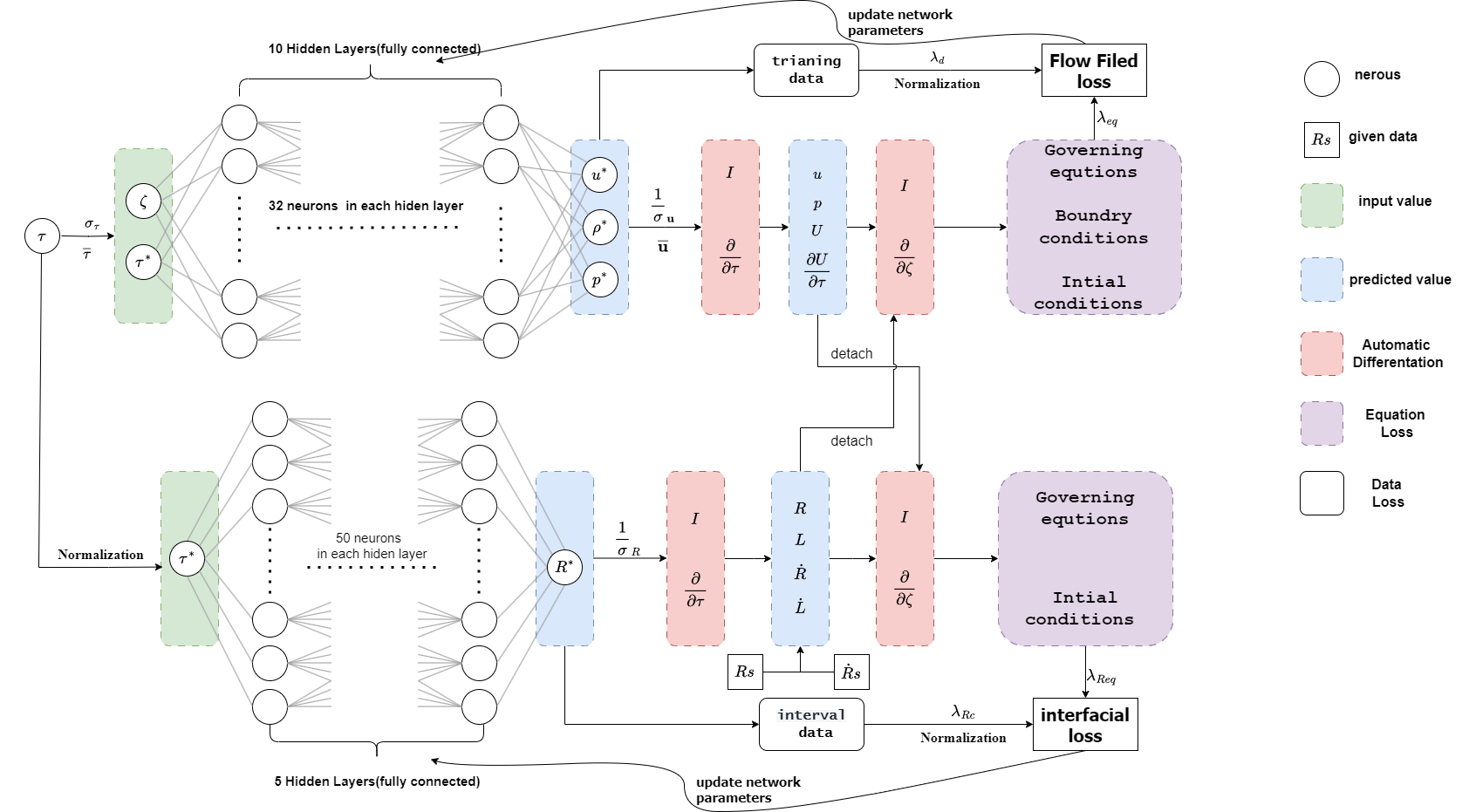} 	
	\caption{Dual-network PINN architecture. \textbf{(Top)} Flow field network with data and physics loss paths. \textbf{(Bottom)} Interface network with interval constraints. Cross-network communication occurs through shared physical constraints in the equation loss calculations.} 
	\label{fig4}
\end{figure}

The coupled nature of underwater explosion (UNDEX) problems demands simultaneous resolution of both the time-dependent gas-liquid interface $R(t)$ and the spatiotemporal flow field $\mathbf{u}(x,t)$. Recognizing their fundamentally different mathematical structures, we propose a dual-network architecture where separate neural networks independently model each component while maintaining physical coupling through shared loss terms as shown in figure \ref{fig4}. 

\textit{Flow Field Network}: The flow field network extends the Case 1 framework, modeling disturbance propagation through a linearly mapped spatial domain $\Omega_m$ with coordinates $(\zeta,\tau)$. The system is governed by left boundary (pressure input determined by interface radius $R(t)$), initial condition (highly compressible state) and right boundary (static water condition).

The solution $\mathbf{u}_m(\zeta,\tau)$ follows an analogous loss formulation to Case 1, ensuring physical consistency while accommodating the moving interface.

\textit{Interface Network}: The interface network solves a multi-material Riemann problem constrained by Euler equations. To address the absence of precise interface data from conventional CFD, we introduce an interval-constrained loss function:
\begin{equation}
L_R(\theta)=\lambda_{Rc} L_{Rc}(\theta)+\lambda_{Req} L_{Req}(\theta)+\lambda_{R0}L_{R0}(\theta)
\end{equation}

The constraint loss penalizes predictions outside CFD-derived bounds:
\begin{equation}
L_{\text{Rc}}(\theta) = \frac{1}{n_{Rc}}\sum_{i=1}^{n_{Rc}} 
\left\|
    \left[\text{ReLU}\left(R_{up}(t_i)-R(t_i)\right)\right]^2 + 
    \left[\text{ReLU}\left(R(t_i)-R_{low}(t_i)\right)\right]^2
\right\|^2
\end{equation}

Equation residuals couple the interface with flow field data:
\begin{align}  
L_{Req1}&= \frac{1}{n_{Req}} \sum_{i=1}^{n_{Req}} \left[
   \frac{\partial \rho_i}{\partial \tau} +
   \frac{\partial}{\partial \zeta} ( \hat{u}(\tau_i)\rho_i ) +
   \rho_i \left( \frac{2u_i}{\zeta_i L_i + R(\tau_i)} + \frac{\dot{L}_i}{L_i} \right)
\right]^2 \\
L_{Req2}&= \frac{1}{n_{Req}} \sum_{i=1}^{n_{Req}} \left[
   \frac{\partial (\rho_i u_i)}{\partial \tau} +
   \frac{\partial}{\partial \zeta} ( \hat{u}(\tau_i)\rho_i u_i + \frac{p_i}{L_i} ) +
   \rho_i u_i \left( \frac{2u_i}{\zeta_i L_i + R(\tau_i)} + \frac{\dot{L}_i}{L_i} \right)
\right]^2 \\
L_{Req3}&= \frac{1}{n_{Req}} \sum_{i=1}^{n_{Req}} \left[
   \frac{\partial E_i}{\partial \tau} +
   \frac{\partial}{\partial \zeta} ( \hat{u}(\tau_i)E_i + \frac{u_i p_i}{L_i} ) +
   \frac{2u_i (E_i + p_i)}{\zeta_i L_i + R(\tau_i)} +
   \frac{E_i \dot{L}_i}{L_i}
\right]^2
\end{align}
where $\hat{u}(\tau)=\frac{1}{L}(u-\zeta \dot R(\tau)+\dot L)$ provides velocity transformation between frames.

\textit{Normalization and Training Strategy}: To ensure numerical stability and accelerate convergence, we implement:

a) Input/Output Normalization:
\begin{equation}
    \tau^{\ast}=\frac{\tau-\overline{\tau}}{\sigma_u},\quad \mathbf{u}^{\ast}=\frac{\mathbf{u}-\overline{\mathbf{u}}}{\sigma_\mathbf{u}},\quad R^{\ast}=\frac{R}{\sigma_R}
\end{equation}

b) Dynamic Weight Scheduling:
\begin{equation}
\lambda_d=0.3+0.6\left(1+\cos{\pi\frac{epoch}{2000}}\right)
\end{equation}
This cosine annealing schedule gradually shifts emphasis from data fidelity ($\lambda_d=0.9$ initially) to physical constraints ($\lambda_{eq}=0.7$ at convergence).

\section{Result and Discussions}\label{sec:4}
\subsection{Training and Performance Evaluation}

\begin{figure}
	\centering 
	\includegraphics[width=0.6\textwidth, angle=0]{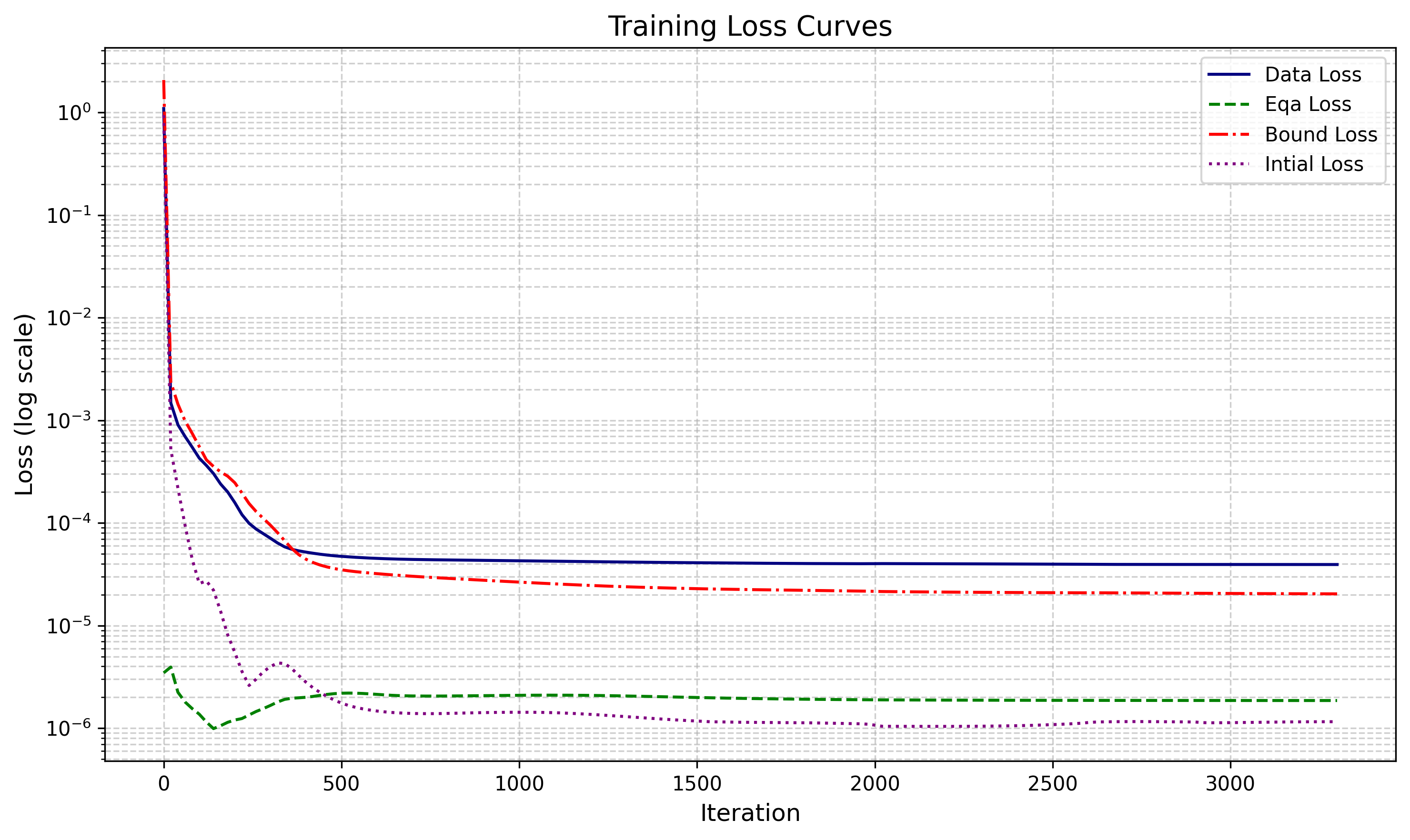}	
	\caption{Training loss evolution for Case Study 1, showing the convergence behavior of: (1) data loss, (2) equation residuals, and (3) initial/boundary condition losses over 3300 training epochs.} 
	\label{fig5}
\end{figure}

The model was trained using an Adam optimizer with learning rate $\eta = 10^{-5}$ and decay rate $\beta = 0.999$. Implemented on an Intel 4070 GPU-equipped system, the training completed in 6798.60 seconds (approximately 1.89 hours) across 3300 epochs, demonstrating stable convergence as shown in figure~\ref{fig5}.

The trained network successfully captured the shockwave propagation dynamics, accurately reconstructing the continuous $u$, $\rho$, and $p$ fields from coarse mesh simulations. Quantitative evaluation yielded L2 error norms: velocity of $3.9782\%$, density of $0.5921\%$ and pressure of $8.8243\%$.

\begin{figure}[htbp]
    \centering
    \begin{subfigure}{\textwidth}
        \centering
          \begin{overpic}[width=0.95\linewidth]{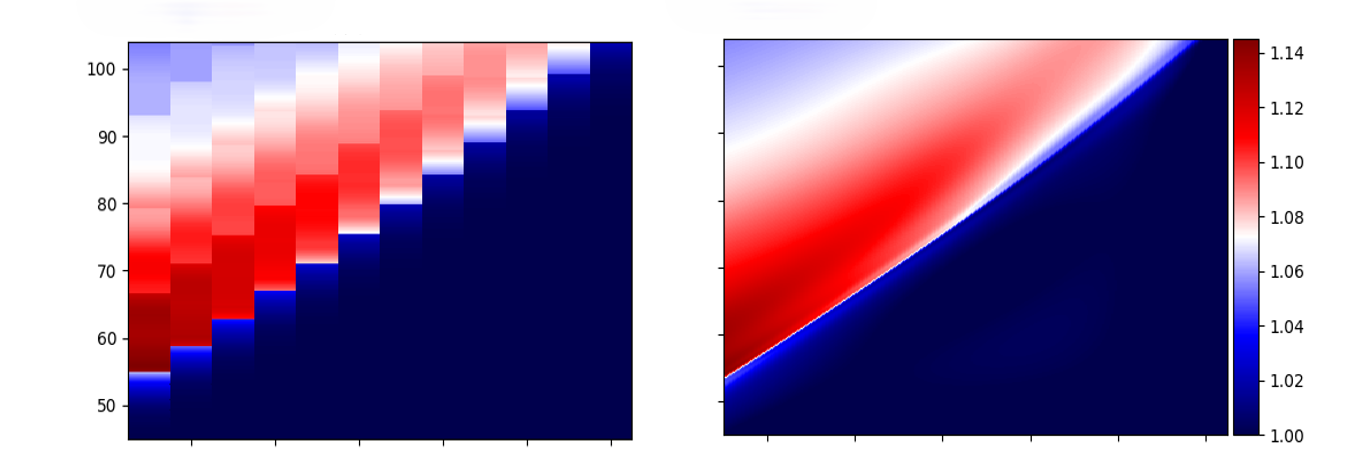}
    \put(0,110){\makebox(0,0)[lb]{\small ($a$)}}
  \end{overpic}
    \end{subfigure}
    \begin{subfigure}{\textwidth}
        \centering
 \begin{overpic}[width=0.95\linewidth]{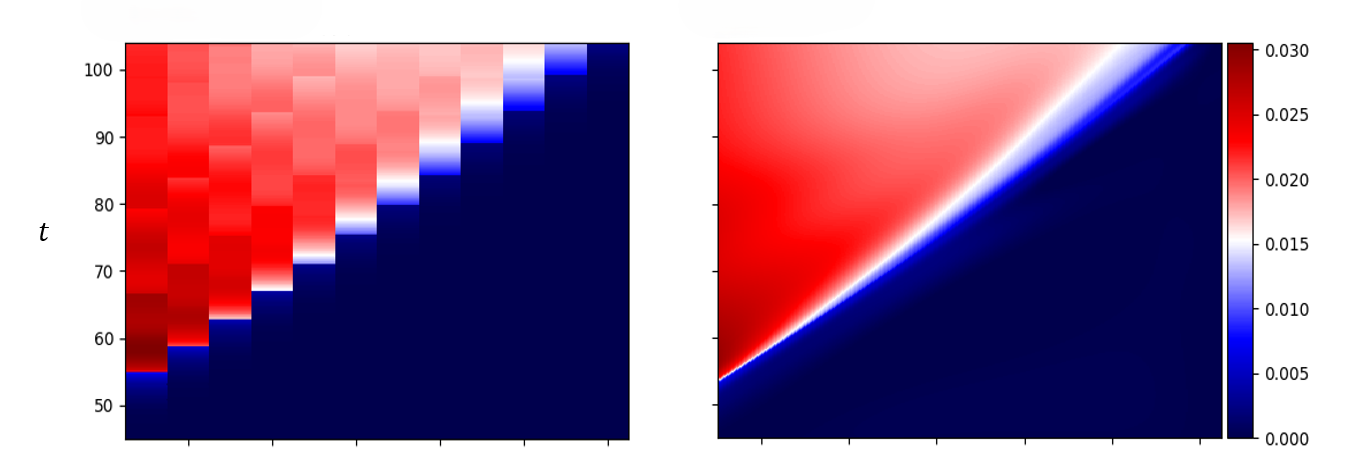}
    \put(0,110){\makebox(0,0)[lb]{\small ($b$)}}
  \end{overpic}
    \end{subfigure}
    
    \begin{subfigure}{\textwidth}
        \centering
  \begin{overpic}[width=0.95\linewidth]{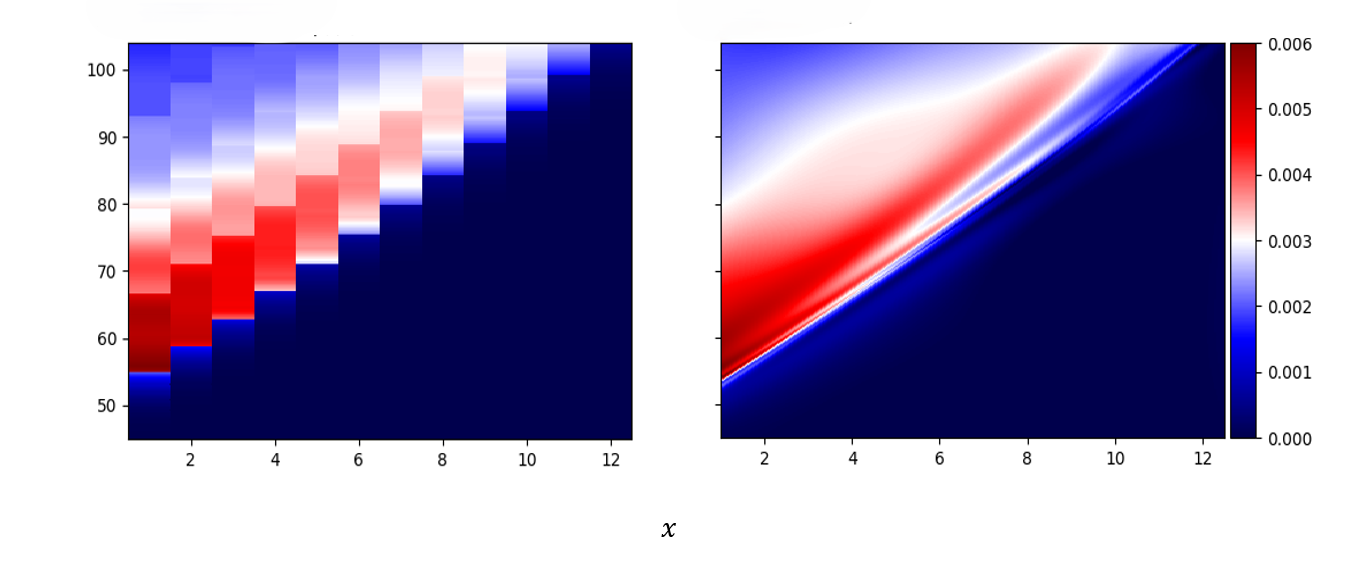}
    \put(0,135){\makebox(0,0)[lb]{\small ($c$)}}
  \end{overpic}
    \end{subfigure}
    \caption{Comparison of CFD reference solutions (left) versus PINN predictions (right) for shock wave propagation: \textbf{($a$)} Density field (\text{g/$cm^3$}); \textbf{($b$)} Velocity field (\text{cm/$\mu$s}); \textbf{($c$)} Pressure field (\si{M\bar}). Spatial coordinates represent distance from detonation point (\si{cm}), with temporal evolution shown from TNT ignition (\si{\mu s}). The superior density and velocity reconstruction accuracy contrasts with slightly higher pressure field errors, attributed to its stronger nonlinearities.}
    \label{fig6}
\end{figure}
As evidenced in figure~\ref{fig6}, the network achieves particularly high fidelity for density and velocity fields, while pressure reconstruction shows marginally larger deviations. This performance difference stems from the pressure field's more abrupt gradients and stronger nonlinearities during shock propagation, which present greater approximation challenges. Visual inspection confirms the network's ability to capture essential physical features while maintaining smoothness across the computational domain.

\subsection{The influence of modifying the constraint conditions on the prediction results}

The dual-network model was trained using consistent optimization parameters across both networks, with learning rate $\eta = 10^{-4}$ and decay rate $\beta = 0.997$. Implemented on an Intel 4070 GPU system, the complete training process required 5952.14 seconds (approximately 1.65 hours), demonstrating stable convergence for both the flow-field and gas-liquid interface networks.

To ensure physically realistic solutions, we developed two progressive constraint formulations for the interface radius predictions. The basic positive constraint enforces strictly positive values through a softplus transformation $R_{\text{base}} = \ln(1 + \exp(R_{\text{raw}}))$, while the enhanced shifted positive constraint $R_{\text{shifted}} = R_{\text{base}} + R_c$ incorporates prior knowledge that the interface must expand beyond the initial TNT charge radius $R_c$. These constraints work synergistically to maintain physical plausibility throughout the prediction domain.

The selection of the initial time parameter $\epsilon = 0.00826\ \si{\micro s}$ represents a critical balance between physical accuracy and numerical stability. This value, corresponding to the first timestep after shockwave emergence, establishes an initial domain length $L = R_c/200$ that properly captures the rapid early-stage dynamics while avoiding the unphysical solutions that can arise from either excessively small or large $\epsilon$ values. Small values may hinder convergence, while large values risk introducing artifacts from the decaying interface velocity profile.

\begin{figure}[htbp]
    \centering 
    \includegraphics[width=\textwidth]{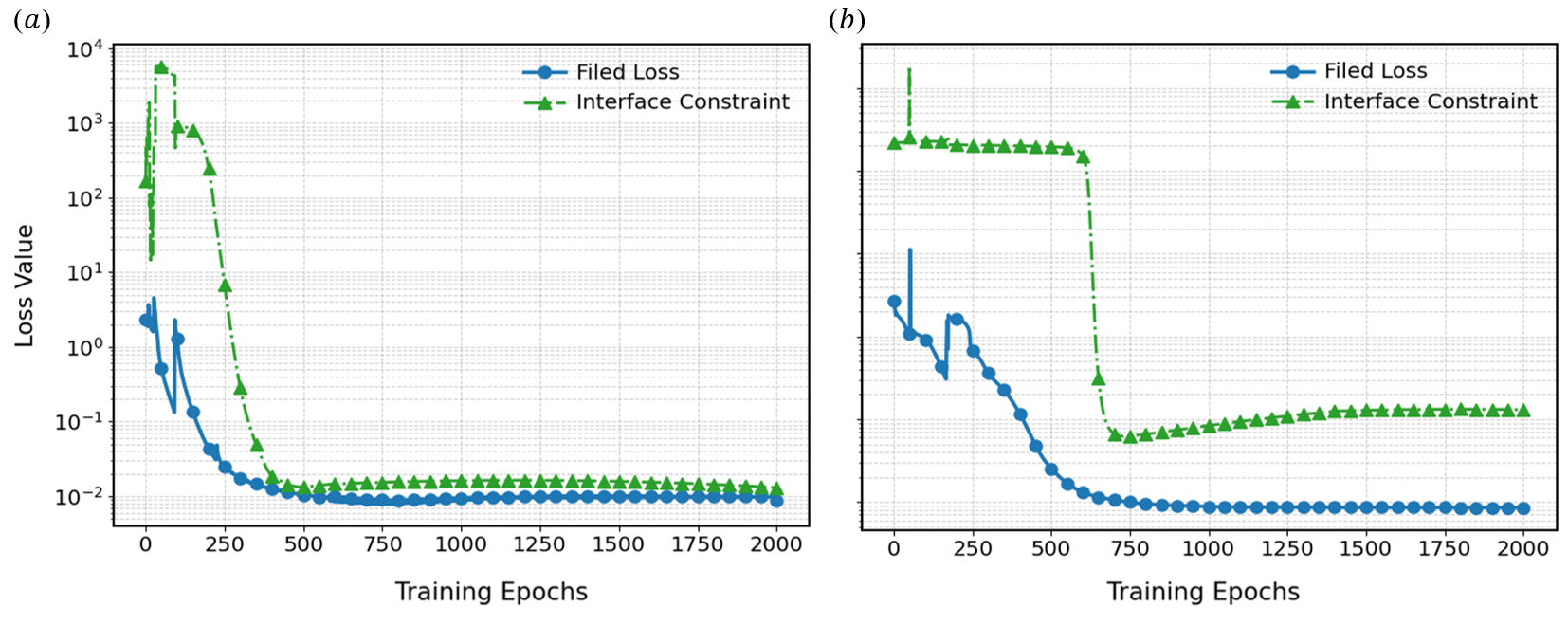}	
    \caption{Training loss evolution across 2000 epochs for Case Study 2:  \textbf{($a$)} basic positive constraint; \textbf{($b$)} shifted positive constraint, showing comparative performance of flow field and interface predictions under different constraint conditions.}
    \label{fig7}
\end{figure}

The training dynamics shown in figure~\ref{fig7} reveal three distinct phases resulting from the strong physical coupling between the gas-liquid interface and flow field networks. During the initial phase, both networks exhibit significant instability as they adapt to the constrained optimization landscape. The intermediate phase shows divergent behavior: while the flow field network begins stable gradient descent, the interface network enters a prolonged plateau period with minimal loss reduction until the flow field approaches convergence. In the final phase, the interface network experiences rapid loss reduction followed by a characteristic rebound before eventual convergence, whereas the flow field network converges monotonically. 

Notably, the shifted positive constraint modifies this progression in two key aspects compared to the basic constraint: (1) the interface network's rapid descent begins around epoch 600, coinciding with the basic constraint's convergence point, and (2) post-convergence fluctuations are more pronounced for the interface network under shifted constraints, while the flow field network remains unaffected. These observations suggest that the shifted constraint's explicit radius offset alters the optimization landscape's topology, particularly for interface predictions.

\begin{figure}[htbp]
	\centering 	\includegraphics[width=\textwidth, angle=0]{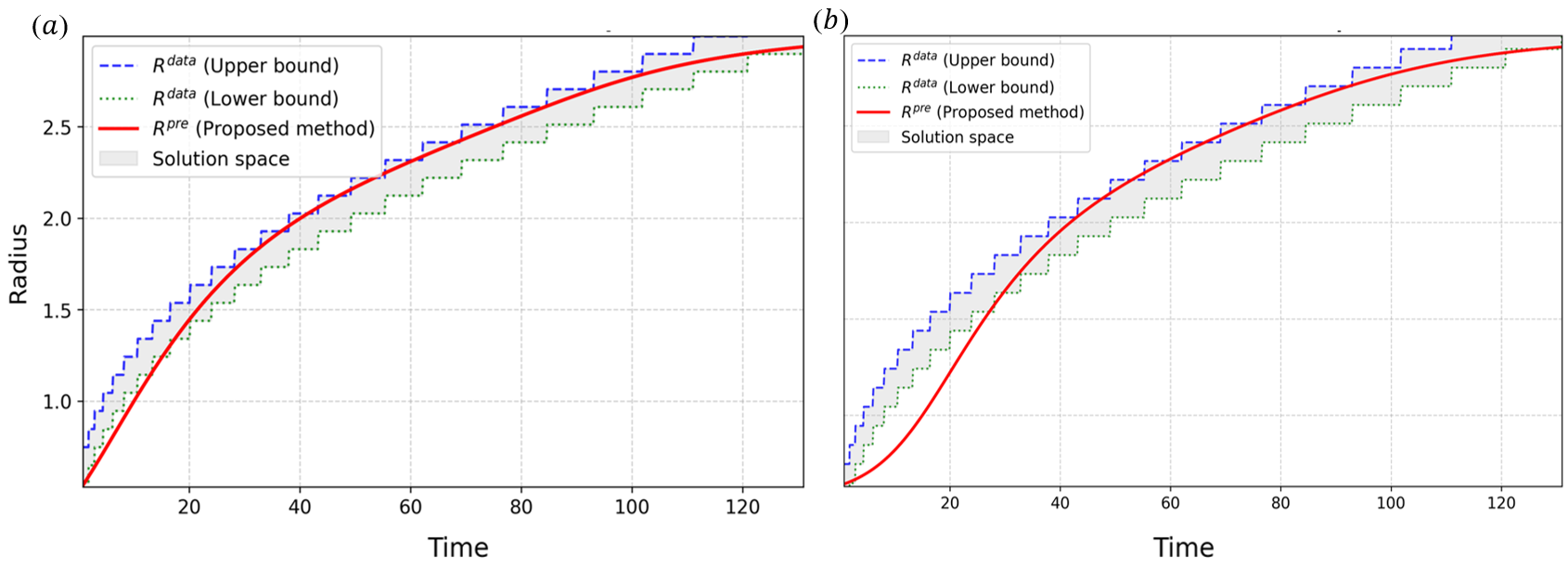}	
	\caption{Comparative evaluation of gas-water interface predictions at epoch 2000:  PINN reconstructions under \textbf{($a$)} basic positive constraint and  \textbf{($b$)} shifted positive constraint versus  CFD-derived interval bounds. The plot demonstrates constraint-dependent prediction accuracy across the temporal domain.} 
	\label{fig8}
\end{figure}

For the case study implementation, we processed approximately 110,000 data points within the triangular region bounded by the gas-liquid interface and shock front $Rs$. The interval constraint arrays were reshaped to match the flow field data geometry, with a 7:3 training-validation split. Figure~\ref{fig8} demonstrates the resulting interface predictions after 2000 epochs of training, comparing PINN outputs against CFD reference solutions for both constraint conditions.

Figure~\ref{fig8} illustrates the discretized interval bounds (blue and green step-like curves) representing the acceptable range for gas-liquid interface predictions. While the basic positive constraint produces results within these bounds that follow the expected trend, the shifted constraint shows early-stage deviations corresponding to its loss function behavior observed during training.

\begin{figure}[h]
    \flushleft
    \includegraphics[width=6in]{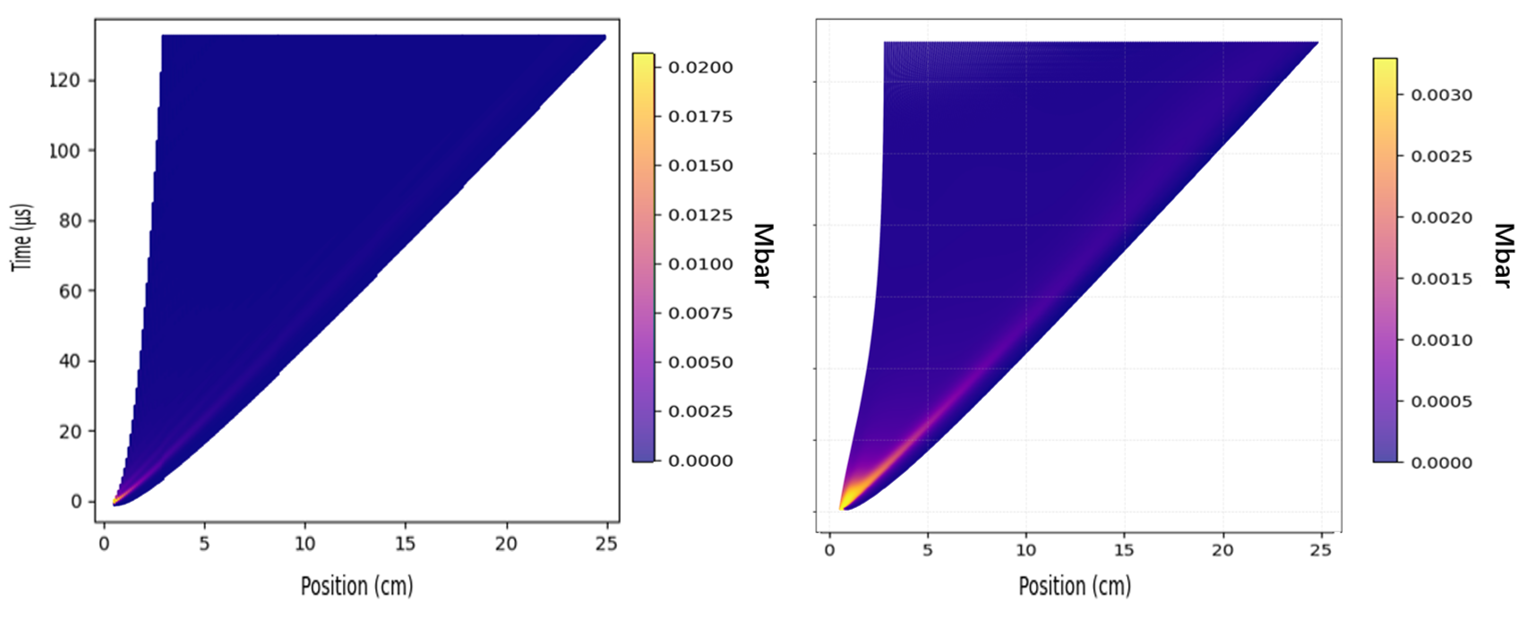}
    \caption{Comparison of pressure field evolution in high-impact water region: (left) CFD reference solutions versus (right) PINN predictions under basic positive constraint. The complete spacetime pressure distribution demonstrates the model's capability in capturing shockwave propagation dynamics.}
    \label{fig9}
\end{figure}

The physical field reconstruction is presented in figure~\ref{fig9}, where we transformed the $\zeta$ coordinate back to physical space $r$. The basic positive constraint captures the general pressure wave propagation pattern, but suffers from significant initial peak pressure discrepancies (L2 error = 14.75\%), rendering it unsuitable for accurate flow field prediction. Given these substantial early-stage errors and the resulting distortion in mid-to-far field propagation, we limit our visualization to the initial explosion phase (0-25 $\mu$s), where the pressure dynamics are most critical.

\begin{figure}[h]
\centering
\begin{subfigure}{\textwidth}
\centering
          \begin{overpic}[width=0.95\linewidth]{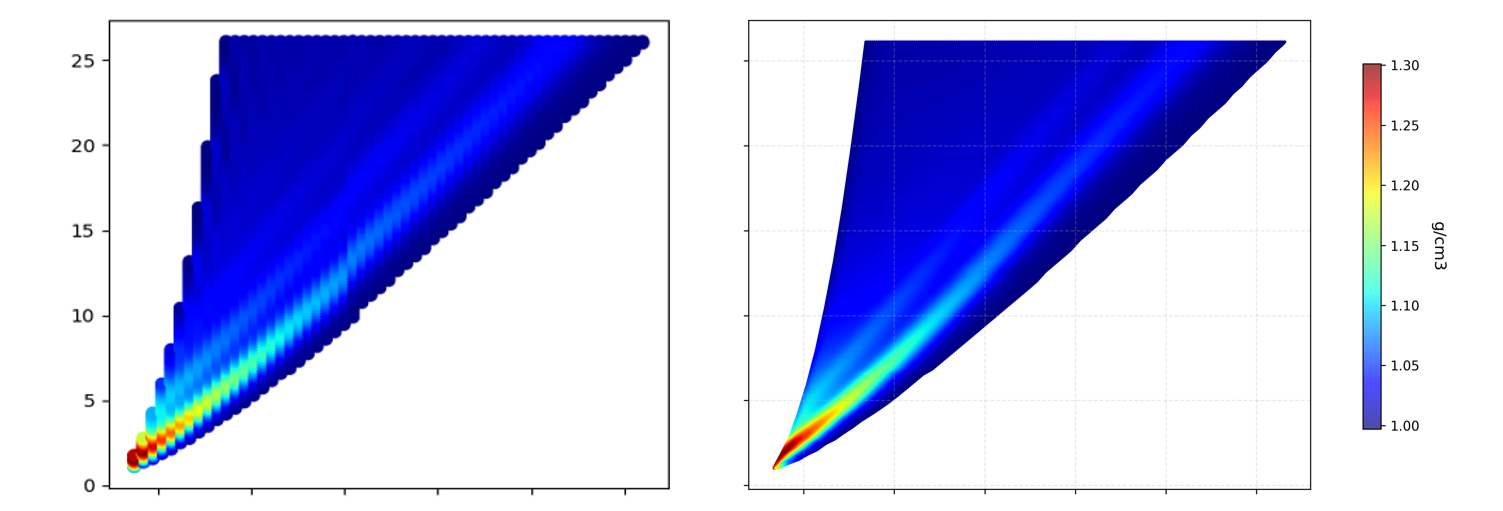}
    \put(0,110){\makebox(0,0)[lb]{\small ($a$)}}
  \end{overpic}
\end{subfigure}

\begin{subfigure}{\textwidth}
\centering
          \begin{overpic}[width=0.95\linewidth]{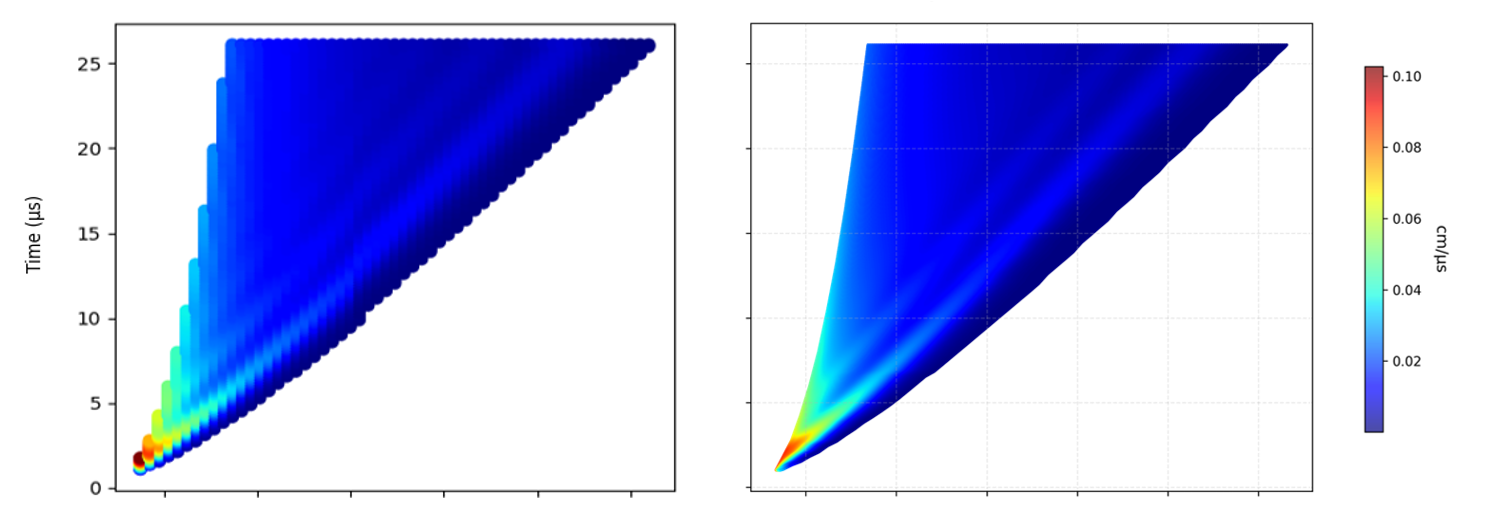}
    \put(0,110){\makebox(0,0)[lb]{\small ($b$)}}
  \end{overpic}
\end{subfigure}

\begin{subfigure}{\textwidth}
\centering
          \begin{overpic}[width=0.95\linewidth]{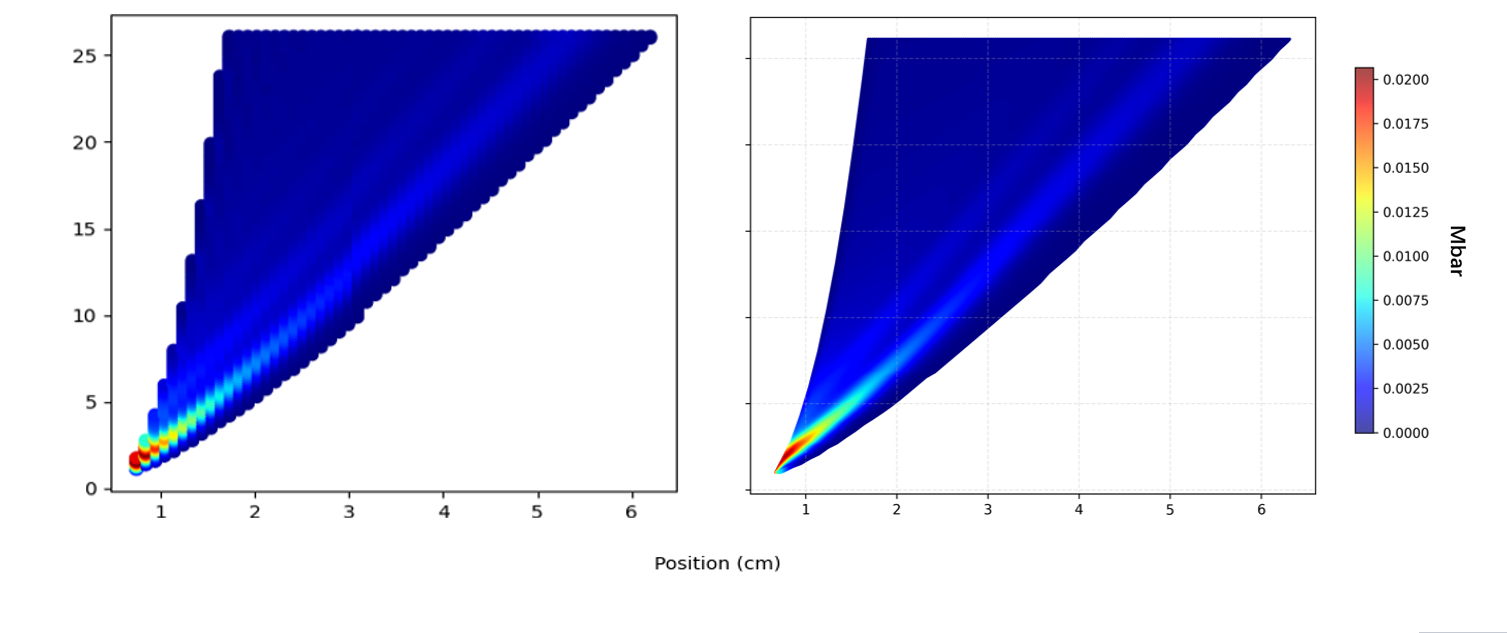}
    \put(0,145){\makebox(0,0)[lb]{\small ($c$)}}
  \end{overpic}
\end{subfigure}

\caption{Comparative analysis of physical fields in the high-impact water region during early explosion (most nonlinear phase):  (left) CFD reference solutions versus (right) PINN predictions under shifted positive constraint with \textbf{($a$)} density, \textbf{($b$)} velocity, and \textbf{($c$)} pressure fields. Spatial coordinates represent distance from explosion center, with temporal zero point ($t=0$) defined at $\epsilon$ after shockwave reaches charge surface $R_c$. Unified color bars enable direct quantitative comparison across all fields.}
\label{fig10}
\end{figure}

Figure~\ref{fig10} presents a comprehensive comparison of density ($\rho$), velocity ($u$), and pressure ($p$) fields between PINN predictions and CFD reference solutions. The results demonstrate strong agreement in both physical patterns and magnitude scales across all three quantities. Quantitative evaluation yields L2 errors of 0.093\% (density), 4.880\% (velocity), and 7.464\% (pressure) on the validation set, confirming the shifted positive constraint's effectiveness in maintaining prediction accuracy within acceptable error bounds for this challenging high-nonlinearity regime.

\begin{figure}[htbp]
\centering
\begin{subfigure}{\textwidth}
\centering
          \begin{overpic}[width=0.6\linewidth]{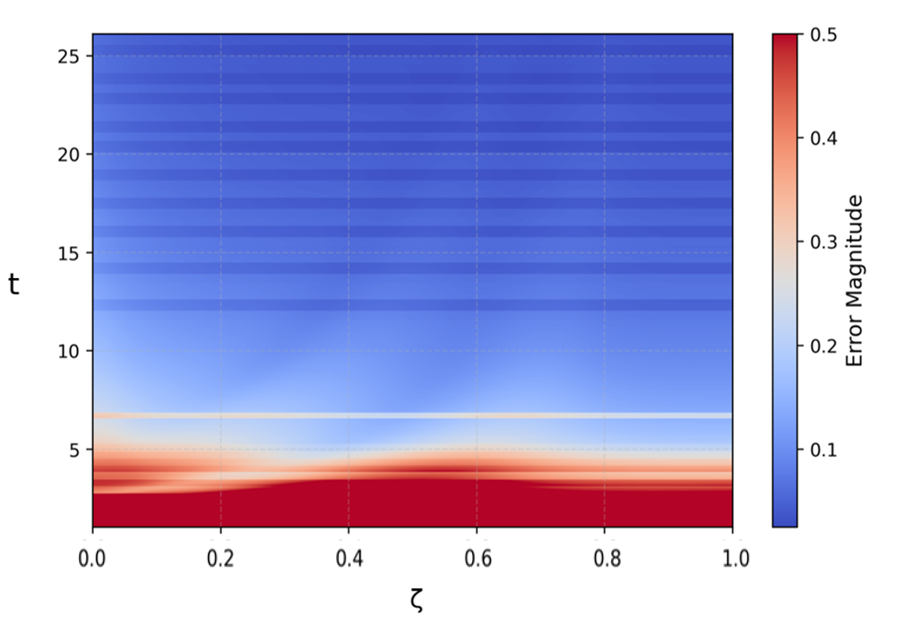}
    \put(0,150){\makebox(0,0)[lb]{\small ($a$)}}
  \end{overpic}
\end{subfigure}

\begin{subfigure}{\textwidth}
\centering
          \begin{overpic}[width=0.95\linewidth]{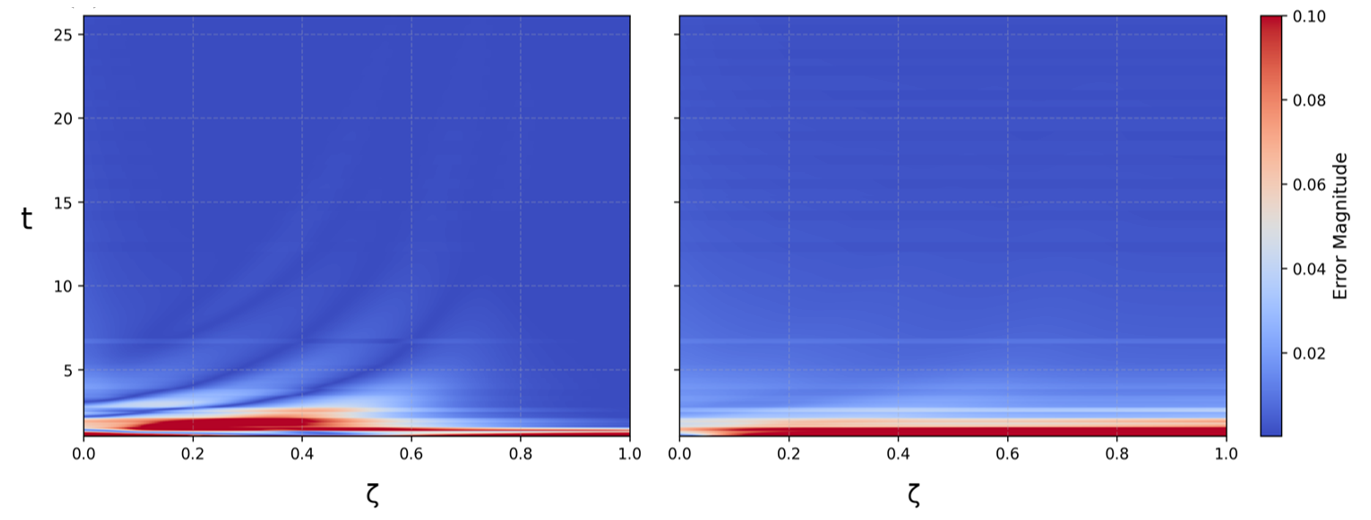}
    \put(0,130){\makebox(0,0)[lb]{\small ($b$)}}
    \put(175,135){\makebox(0,0)[lb]{\small ($c$)}}
  \end{overpic}
\end{subfigure}
\caption{Euler equation residuals under shifted positivity constraint showing: (\textbf{$a$}) continuity, (\textbf{$b$}) momentum, and (\textbf{$c$}) energy equation deviations. The color scale represents absolute residual values between equation sides.}
\label{fig11}
\end{figure}

Figure~\ref{fig11} presents the residual distributions of the three conservation equations across the computational domain, calculated as absolute differences between equation sides. Key observations reveal that maximum residuals concentrate in early timesteps, with continuity equation residuals substantially exceeding those of momentum and energy equations. This temporal and equation-specific pattern suggests particular challenges in modeling initial shock formation dynamics.

Our comparative analysis yields two principal findings. First, the basic positive constraint demonstrates superior accuracy for gas-liquid interface prediction, while the shifted constraint better captures flow field dynamics. We attribute this performance difference to inherent limitations in early-stage CFD data accuracy, which may deviate from theoretical expectations. Second, the constraint formulation significantly influences network convergence behavior. While each network naturally converges independently during loss minimization, their physical coupling creates interdependent convergence patterns. When objective errors prevent simultaneous minimization, the more predictive network continues converging while the weaker one (particularly the shifted-constraint interface network) exhibits compromised performance. This phenomenon underscores the importance of balanced constraint design for coupled physical systems.
\section{Validation}\label{sec:5}
\begin{figure}[htbp]
\centering
\begin{subfigure}{\textwidth}
    \centering
    \includegraphics[width=0.8\textwidth]{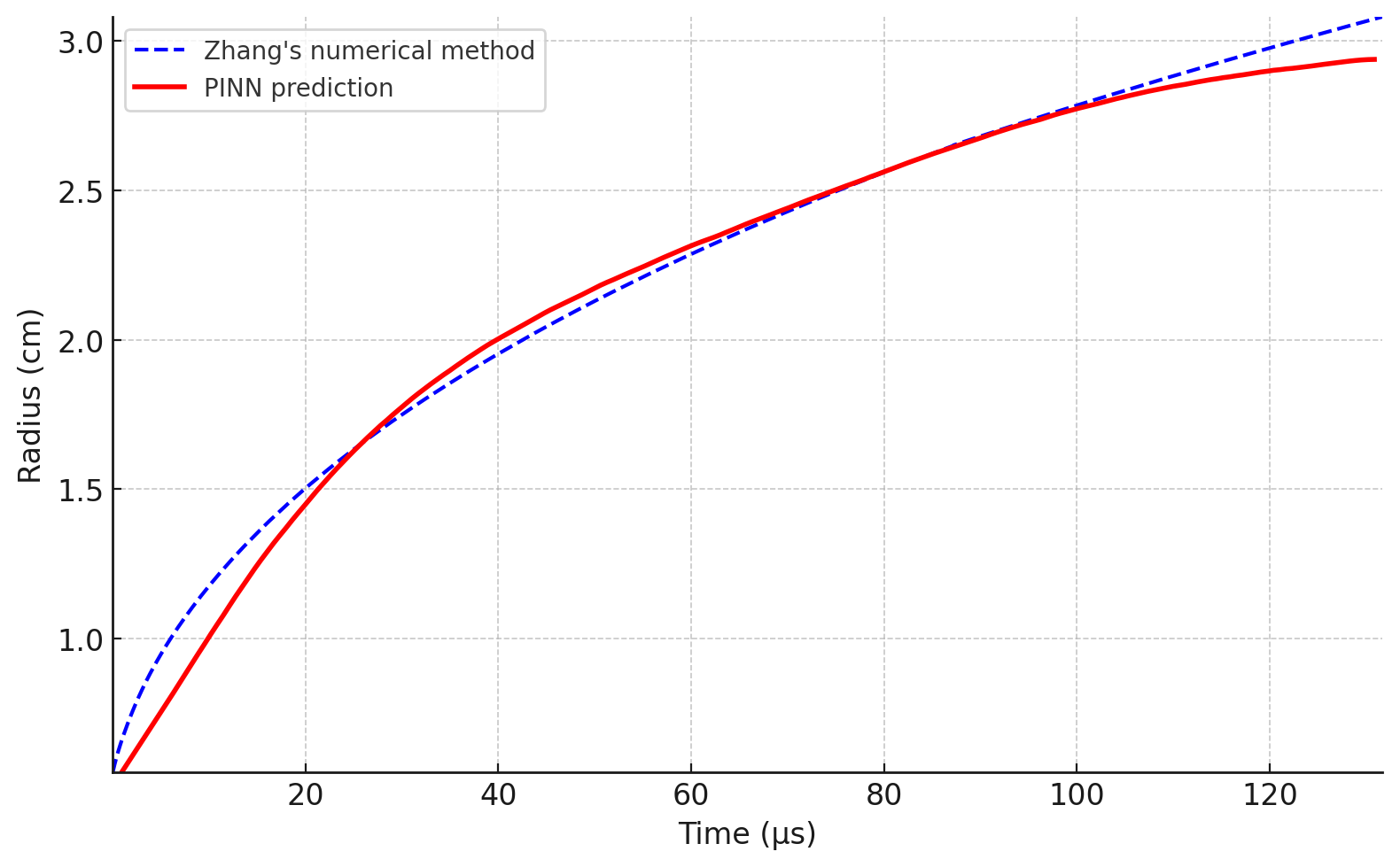}

\end{subfigure}
    \caption{Comparison of gas–liquid interface prediction. The PINN-predicted interface radius evolution (red solid line) is compared against Zhang's numerical results (blue dashed line) under a TNT-equivalent charge of 1.1\,g. }
\label{fig12}
\end{figure}

In this study, we validate the effectiveness of PINN in predicting gas–liquid interface evolution using an open-source solver based on Unified Bubble Theory, which numerically solves the Euler equations in a scaled coordinate system using the Discontinuous Galerkin method. This solver has been extensively validated against both theoretical and experimental data, making it a reliable reference \citep{zhang2023unified,li2023experimental}. For validation, we set the solver parameters as follows: TNT equivalent of 1.1 g, water depth at 0 m, and material properties of water and TNT adjusted according to table~\ref{tab:water_tnt_compact}. The gas–liquid interface positions predicted by the PINN interface network are directly compared with numerical results obtained from the solver, as shown in figure~\ref{fig12}. While the interface predictions show satisfactory agreement in the mid-time range (20$\mu$s to 100$\mu$s), notable deviations are observed during the initial (0–20$\mu$s) and final (100–130$\mu$s) phases. To improve the model’s fidelity, future extensions will incorporate refined modeling of the internal TNT gas state, as well as surface tension and water viscosity effects.

\begin{figure}[htbp]

\begin{subfigure}{\textwidth}
    \centering
    \includegraphics[width=0.8\textwidth]{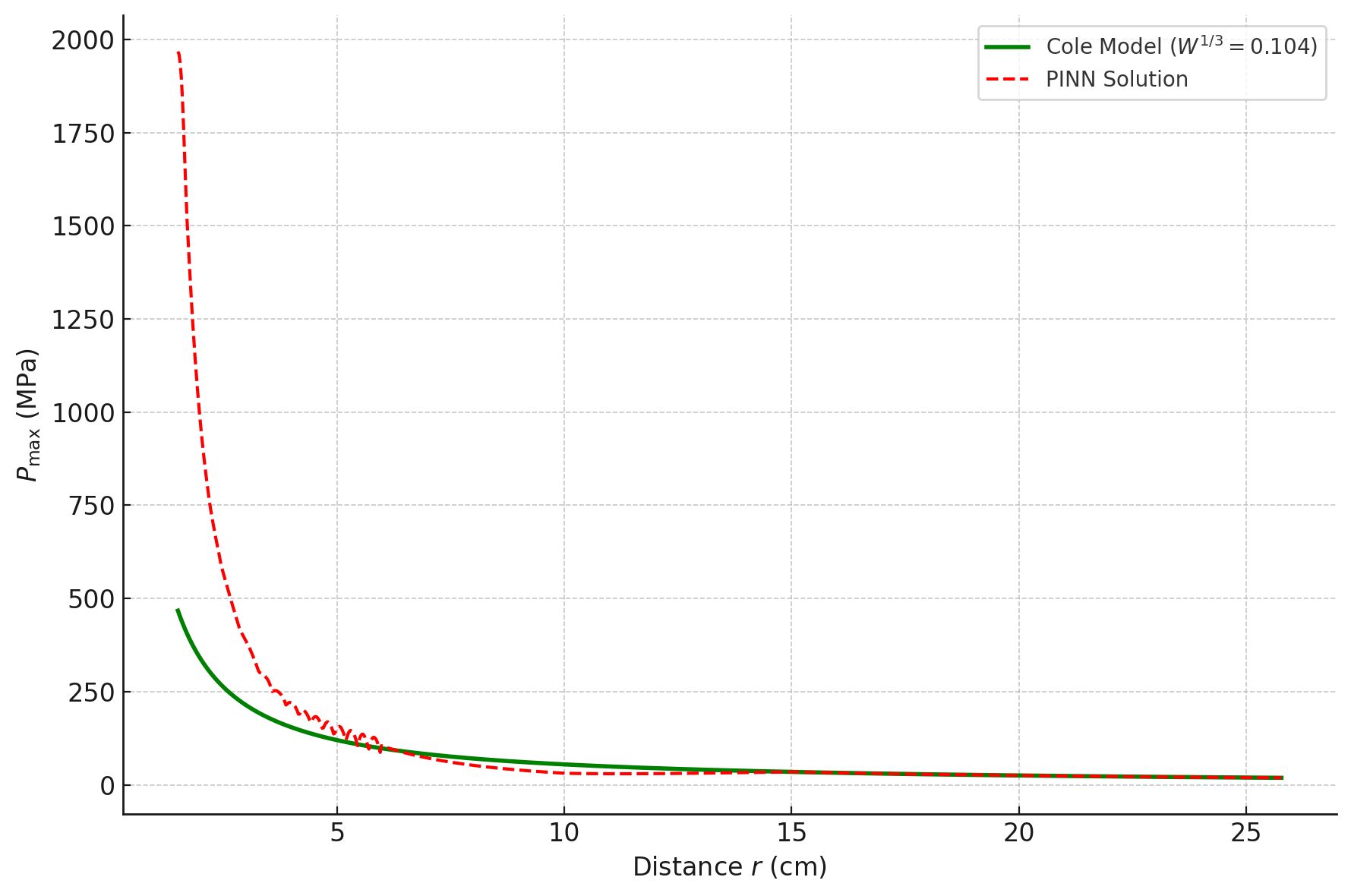}

\end{subfigure}
    \caption{Comparison of peak pressure profiles. The maximum pressure $P_{\text{max}}$ predicted by PINN (red dashed line) is compared with the classical Cole model (green solid line, $W^{1/3} = 0.104$). }
\label{fig13}
\end{figure}

Additionally, to assess the validity of the PINN-predicted peak pressure profiles, we compare them with the classical Cole empirical model \citep{cole1948underwater}, which has long served as a standard reference for underwater explosion pressure analysis. The Cole model has been extensively validated through both experimental and numerical studies \citep{moon2017assessment,nowak2023small}, and is particularly reliable in regions located beyond ten times the TNT charge radius from the explosion center\citep{OUYANG2025113567,jha2014under}. The empirical formula for peak pressure is given by:

\begin{equation}
P_{max} = 52.39 \left(\frac{W^{1/3}}{r}\right)^{1.13} \quad (\text{MPa}),
\end{equation}
where $W$ is the TNT equivalent in kg, and $r$ is the distance from the explosion center in meters. The PINN-predicted peak pressures are obtained by traversing the pressure field to track the attenuation characteristics of the peak pressure with distance. As shown in figure~\ref{fig13}, our PINN predictions agree well with the Cole empirical model at distances beyond 5 cm from the explosion center. This strong consistency in the mid-to-far field highlights the significant potential of the PINN framework for reliable underwater explosion assessment.

\section{Conclusion}\label{sec:6}
This study demonstrates that physics-informed neural networks (PINNs) effectively model nonlinear underwater explosion dynamics across initial phase and far-field regimes. For far-field shockwave propagation, PINNs successfully reconstructed continuous physical fields from sparse data. In initial phase scenarios, our novel interval-constrained approach enabled accurate gas-liquid interface prediction by combining weak supervision with physical constraints. Comparative analysis revealed the shifted positive constraint's superiority for flow field prediction in shock-dominated regions, while the basic constraint provided better interface positioning due to its stronger network regularization. The framework successfully generated continuous 1D representations of both flow fields and interfaces by integrating theoretical models with simulation data. These results establish PINNs as a promising tool for underwater explosion modeling, overcoming traditional limitations in handling moving boundaries and strong discontinuities. Future work will enhance model generalizability and accuracy, particularly for multi-dimensional scenarios, further advancing PINN applications in scientific and engineering domains where physical consistency and computational efficiency are paramount.

\begin{bmhead}[Funding.]
This work is supported by the developing Project of Science and Technology of Jilin Province (20240402042GH).
\end{bmhead}

\begin{bmhead}[Declaration of interests.]
The authors report no conflict of interest.
\end{bmhead}

\begin{bmhead}[Author ORCIDs. ]\\
\orcidlink{0009-0007-4484-516X}Fulin  Xing:\href{https://orcid.org/0009-0007-4484-516X}{https://orcid.org/0009-0007-4484-516X}\\
\orcidlink{0009-0007-0797-6822}Junjie Li: \href{https://orcid.org/0009-0007-0797-6822}{https://orcid.org/0009-0007-0797-6822}   \\
\orcidlink{0009-0004-0202-3641}Ze Tao:\href{https://orcid.org/0009-0004-0202-3641}{https://orcid.org/0009-0004-0202-3641}\\
\orcidlink{0000-0002-8573-450X}Fujun Liu:\href{https://orcid.org/0000-0002-8573-450X}{https://orcid.org/0000-0002-8573-450X}
\end{bmhead}

\begin{bmhead}[Author contributions.]
F.X. Calculation, data analyzing and manuscript writing. J.L. and Z.T. advised on the physical modeling and computational methods.  F.L. Supervision and Resources. Y.T. Supervision. All authors contributed to the
 paper writing and English editing. All authors have contributed to the manuscript.
\end{bmhead}

\begin{bmhead}[Data availability. ]
Data will be made available on request.
\end{bmhead}

\begin{appen}

\section{Linear mapping of the spherically symmetric Euler equations}\label{appA}
After performing a linear mapping on $r$, we attempted to construct a coordinate system based on $\zeta$ and $t$, but found that $\zeta$ is explicitly dependent on time $t$. In other words, $t$ and $\zeta$ are not independent. Therefore, when taking partial derivatives, in order to ensure that the other variable is held constant, we have:
\begin{equation}
    r(t)=\zeta L(t)+R(t), \quad \left. \frac{\partial r}{\partial t} \right|_{\zeta} = \zeta \dot{L} + \dot{R}.  \label{A1}
\end{equation}
\begin{equation}
    \left. \frac{\partial}{\partial t}\, f\bigl(r(t),\,t\bigr) \right|_{\zeta}
   \;=\;
   \frac{\partial f}{\partial t}
   \;+\;
   \frac{\partial f}{\partial r}\,
   \frac{\partial r}{\partial t}
   \label{A2}
\end{equation}
In conjunction with equations \eqref{A1} and \eqref{A2}, we define the partial derivative with respect to $t$ at fixed 
$\zeta$ as:
\begin{equation}
    \frac{\hat{\partial} \rho}{\hat{\partial} t} = \frac{\partial \rho}{\partial t} + \frac{\partial \rho}{\partial r} \left( \zeta \dot{L} + \dot{R} \right)
\end{equation}
and:
\begin{equation}
    \frac{\partial}{\partial r} = \frac{1}{L} \frac{\partial}{\partial \zeta}.
\end{equation}
We now proceed to map the continuity equation.where the unsteady term is given by:
\begin{equation}
    \frac{\partial \rho}{\partial t} = \frac{\hat{\partial} \rho}{\hat{\partial} t} - \frac{\zeta \dot{L} + \dot{R}}{L} \frac{\partial \rho}{\partial \zeta},\label{A5}
\end{equation}
 and the convective term and the axisymmetric source term are given by:
\begin{equation}
    \frac{\partial (\rho u)}{\partial r} = \frac{1}{L} \frac{\partial (\rho u)}{\partial \zeta}\label{A6}
\end{equation}
\begin{equation}
    -\frac{2u}{r} \rho = -\frac{2\rho}{\zeta L + R} u \label{A7}
\end{equation}
By rearranging equations \eqref{A5}, \eqref{A6}, and \eqref{A7}, the continuity equation under linear mapping can be obtained as:
\begin{equation}
    \frac{\hat\partial \rho}{\hat\partial t} + \frac{\partial}{\partial \zeta} \left( \frac{\rho (u - \dot{R} - \zeta \dot{L})}{L} \right) = -\frac{2u}{\zeta L + R} \rho-\frac{\rho \dot L}{L}
\end{equation}
Similarly, the momentum conservation equation under linear mapping take the following forms:
\begin{equation}
    \frac{\hat\partial \rho u}{\hat\partial t} + \frac{\partial}{\partial \zeta} \left( \frac{\rho u (u - \dot{R} - \zeta \dot{L})+p}{L} \right) =\rho u \left(-\frac{2u}{\zeta L + R} -\frac{\dot L}{L}\right)
\end{equation}
 To simplify the equations, we introduce the velocity $\hat{u}$ in the $\zeta$-coordinate system:
\begin{equation}
    \hat{u} = \frac{d\zeta}{dt} = \frac{1}{L} \left( u - \zeta \dot{L} - \dot{R} \right)
\end{equation}
Thus, the final form under linear mapping is obtained, which we now compare with the original form of the one-dimensional spherical Euler equations:

\begin{equation}
\left\{
\begin{array}{l}
\displaystyle
\frac{\partial \rho}{\partial t} + \frac{\partial (\rho u)}{\partial r} = -\frac{2u}{r}\rho \\
\displaystyle
\frac{\partial (\rho u)}{\partial t} + \frac{\partial (\rho u^2 + p)}{\partial r} = -\frac{2u^2}{r}\rho
\end{array}
\right.
\quad \Longleftrightarrow \quad
\left\{
\begin{array}{l}
\displaystyle
\frac{\hat\partial \rho}{\hat\partial t} + \frac{\partial (\rho\hat{u})}{\partial \zeta}  
=  \frac{-2u}{r(t)}\rho - \frac{\dot{L}}{L}\rho  \\[1ex]
\displaystyle
\frac{\hat\partial (\rho u)}{\hat\partial t} + \frac{\partial \left( \hat{u}\rho u + \frac{p}{L} \right) }{\partial \zeta}
= \frac{-2u^2}{r(t)}\rho - \frac{\dot{L}}{L}\rho u 
\end{array}
\right.
\end{equation}
It can be seen that, after linear mapping, the geometric source term in the one-dimensional spherical Euler equations includes an additional term, which appears as the product of $\frac{\dot{L}}{L}$ and the unsteady term. This arises because a control volume in $\zeta$-space that satisfies conservation laws will undergo temporal expansion in the $r$-space, leading to additional inflow. This term can be explicitly embedded into the physical constraints, highlighting the remarkable flexibility and universality of PINN in physical computations.

\end{appen}\clearpage

\bibliographystyle{jfm}
\bibliography{jfm}


\end{document}